\documentclass[11pt,preprint,superscriptaddress,aps,showkeys,nofootinbib]{revtex4}

\usepackage{amssymb,amsmath,amsfonts}
\usepackage{graphicx}
\usepackage{graphics}
\usepackage{eepic,epsfig}
\usepackage{indentfirst}
\usepackage[utf8]{inputenc}
\usepackage[T1]{fontenc}
\usepackage{amsmath}
\usepackage{subfigure}
\usepackage{hyperref}

1.60cm \makeatletter \@addtoreset{equation}{section}

\makeatother

\begin{document}

\title{Scalar Casimir Effects in a Lorentz Violation Scenario Induced by the Presence of Constant Vectors}

\author{E. R. Bezerra de Mello}
\email{emello@fisica.ufpb.br}
\affiliation{Departamento de Física, Universidade Federal da Paraíba \\ Caixa Postal 5008, 58051-970, João Pessoa, Paraíba, Brazil}

\author{M. B. Cruz}
\email{messiasdebritocruz@gmail.com}
\affiliation{Centro de Ciências Exatas e Sociais Aplicadas, Universidade Estadual da Paraíba \\ CEP 58706-550, Patos, Paraíba, Brazil}


\begin{abstract}

In this work, we consider a theoretical model that presents a violation of Lorentz symmetry in the approach of quantum field theory. The theoretical model adopted consists of a real massive scalar quantum field confined in the region between two large parallel plates. The violation of Lorentz symmetry is introduced by a CPT-even, aether-like approach, considering a direct coupling between the derivative of the scalar field with two orthogonal constant vectors. The main objective of this paper is to analyze the modification of the Casimir energy and pressure caused by the anisotropy of space-time as a consequence of these couplings. The confinement of the scalar quantum field between the plates is implemented by imposing boundary conditions on them. 
\end{abstract}


\maketitle


\section{Introduction}
\maketitle

In 1948, H. B. Casimir proposed a theoretical model to investigate the structure of the quantum vacuum associated with electromagnetic fields \cite{Casimir:1948dh}. Specifically, he assumed that the presence of two large, parallel, and isolated plates could be considered for this investigation. From a classical point of view, the vector wave number parallel to the plates suffers no restrictions; however, the vector orthogonal to them must be discretized. In his original work, Casimir predicted that the quantum fluctuations of the electromagnetic field also satisfy these conditions. By subtracting the zero-point energy for photons in the presence and absence of two parallel flat plates, he obtained the result that the plates attract each other with a force per unit area given by:
\begin{equation}
	\frac{F}{A} = - \frac{\pi^2 \hbar c}{240 a^4}\   ,
\end{equation}
where $A$ is the area of the plates and $a$ is the distance between them. This effect was experimentally observed ten years later by M. J. Sparnaay \cite{Sparnaay:1958wg}; however, only in the 1990s were experiments able to confirm the Casimir effect with a high degree of accuracy \cite{Lamoureux}. This effect is one of the most direct manifestations of the existence of vacuum quantum fluctuations. Initially little studied, from the 1970s, the scientific community from several areas began to explore the phenomenon, especially in Quantum Field Theory (QFT).

In general, we can say that the Casimir effect is a consequence of boundary conditions imposed on quantum fields. These boundaries can be material media, interfaces between two phases of the vacuum, or even space-time topologies. The simplest way to theoretically study the Casimir effect is through the presence of two parallel plates placed in a vacuum.  However, other situations have been studied. For example, the analysis of the Casimir force for a piston configuration in $R^3$, with one dimension being slightly curved and the other two being infinite, was performed in \cite{Oikonomou:2009se}. Moreover, the Casimir energy associated with massless Majorana and bosonic vector fields confined between two parallel plates has been investigated in \cite{Oikonomou:2009zr}.

Since QFT is based on the theory of Special Relativity, Lorentz symmetry is generally assumed to be preserved. However, there are other theories that propose models where this symmetry is violated \cite{Horava}, such as the Horava-Lifshitz (H-L) model, which introduces an anisotropy between space and time. Consequently, the space-time anisotropy in a given QFT model can modify the spectrum of the Hamiltonian operator. The violation of Lorentz symmetry (LV) has also been questioned under both theoretical and experimental contexts. V. A. Kostelecky and S. Samuel \cite{Kostelecky:1988zi} presented a mechanism in string theory that allows for the violation of Lorentz symmetry at the Planck energy scale. This mechanism results in a non-vanishing vacuum expectation value of some vector or tensor components, which implies preferential directions in space-time. Other possible mechanisms for the violation of Lorentz symmetry include space-time non-commutativity \cite{Carroll:2001ws, Anisimov:2001zc, Carlson:2001sw, Hewett:2000zp, Bertolami:2003nm}, variation of coupling constants \cite{Kostelecky:2002ca, Anchordoqui:2003ij, Bertolami:1997iy}, and modifications of quantum gravity \cite{Alfaro:1999wd, Alfaro:2001rb}.

As mentioned earlier, the violation of Lorentz symmetry has been under question since the results obtained in \cite{Kostelecky:1988zi}. If we accept this violation, an immediate question arises: What is the influence of this violation on a given theoretical model from a quantum point of view? It is expected that this violation will certainly affect the entire structure of the theory. One of the most important consequences of the violation concerns the energy spectrum of the modified Hamiltonian operator. In fact, a QFT model is physically sensible only if it possesses a vacuum state. Therefore, an important question to be answered is: How is the vacuum structure modified by the presence of Lorentz symmetry violation? This question cannot be answered solely on theoretical grounds and requires experimental verification. Although the violation of Lorentz symmetry occurs at large-scale energy, some remnants of this violation can be present in low-energy phenomena \cite{Liberati}. Due to the high precision in the measurement of the Casimir force, the corresponding experiment is a good candidate to detect the remnants left by Lorentz symmetry breaking.

In this sense, the study of Lorentz symmetry violation has become of great theoretical and experimental interest. Due to the high precision in the measurement of the Casimir force, the corresponding experiment can be used as a laboratory to study possible traces left by the LV. The first studies on the Casimir effect in the context of Lorentz-violating theories were conducted in \cite{FrankTuran, Escobar} for different extensions of the Quantum Electrodynamics (QED) that break Lorentz symmetry. An analysis of the Casimir effect associated with massless scalar and fermionic quantum fields confined between two large plates, considering space-time anisotropy in the Horava-Lifshitz formalism, was developed in \cite{Ulion:2015kjx} and \cite{Deivid}, respectively. Following the same line of investigation, more recently, the Casimir effect associated with a massive real scalar field was studied in \cite{Maluf2020}. The analysis of Casimir effects associated with massive real scalar and fermionic fields, considering CPT-even Lorentz symmetry breaking in an aether-like scenario through the direct coupling between the field's derivative and an arbitrary constant four-vector, was investigated in \cite{Messias2017} and \cite{Messias2019}, respectively. The analysis of the Casimir energy and topological mass associated with a massive scalar field in the LV scenario was considered in \cite{Messias2020}. In \cite{Erdas2020}, the influence of the presence of a constant magnetic field on the Casimir effect in the Lorentz-violating scalar field was considered. Moreover, the thermal effect on the Casimir energy and pressure caused by the Lorentz-violating scalar field was investigated in \cite{Messias2018}.

In this paper, we aim to continue the investigation carried out in the papers mentioned in the preceding paragraph. However, we will now consider the presence of two orthogonal constant vectors that interact with a real scalar quantum field confined between two large parallel plates. In this context, we will impose specific boundary conditions on the fields that should be satisfied under the plates. This analysis can be seen as a contribution to the existing literature. The presence of these two orthogonal constant vectors will introduce Lorentz violation, making the analysis of the Casimir effects more complex.

This work is organized as follows. In Section \ref{Model}, we provide a brief description of the theoretical model for the real scalar field with terms responsible for space-time anisotropy. In Section \ref{Casimir_LV}, we focus on calculating the Casimir energies and pressures while considering three different boundary conditions obeyed by the quantum fields on the plates. In Section \ref{Conclusions}, we summarize our most important results. We will adopt natural units, $\hbar=c=1$, and a metric signature of $-2$


\section{Model Setup}\label{Model}

In this section, we will introduce the theoretical model that we want to analyze. This model is composed of a massive real scalar quantum field whose dynamics are governed by the lagrangian density below:
\begin{eqnarray}
\label{lagrangedensity}
{\cal{L}}=\frac{1}{2}\left[\eta^{\mu\nu}\partial_\mu\phi\partial_\nu\phi+{\lambda_1} (u\cdot\partial\phi)^2+{\lambda_2} (v\cdot\partial\phi)^2-\mu^2\phi^2\right] \ .
\end{eqnarray} 
The violation of Lorentz symmetry is represented by the presence of two arbitrary constant vectors $u^\mu$ and $v^\mu$ coupled to the scalar field through its derivative. The dimensionless parameters $\lambda_i$, for $i=1, \ 2$, are considered much smaller than unity in this analysis and are introduced to investigate the influence of LV on the physical results obtained in this paper. Moreover, the quantity $\mu$ in Eq. \eqref{lagrangedensity} is defined in terms of the mass in natural units ($\mu=m c/\hbar$).

In this formalism, the modified Klein-Gordon equation (KG) reads,
\begin{eqnarray}
	\label{KG_equation}
\left[\Box+\lambda_1(u\cdot\partial)^2+\lambda_2(v\cdot\partial)^2+\mu^2\right]\phi(x)=0 \ .
\end{eqnarray}
The energy-momentum tensor can be obtained from Eq. \eqref{lagrangedensity} using the standard definition, which is given by:
\begin{eqnarray}
T^{\mu\nu}&=&\frac{\partial{\cal{L}}}{\partial(\partial_\mu\phi)}(\partial^\nu\phi) -\eta^{\mu\nu}{\cal{L}} \ . 
\end{eqnarray}
So, it reads,
\begin{eqnarray}
\label{energy-tensor}
T^{\mu\nu}&=&(\partial^\mu\phi)(\partial^\nu\phi)+\lambda_1 u^\mu(\partial^\nu\phi)(u\cdot\partial\phi)+\lambda_2 v^\mu(\partial^\nu\phi) (v\cdot\partial\phi)-\eta^{\mu\nu}{\cal{L}} \  .
\end{eqnarray}
Due to the presence of LV terms,  the energy-momentum tensor is not symmetric. Its antisymmetric part is given by,
\begin{eqnarray}
	T^{\mu\nu}-T^{\nu\mu}=\lambda_1\left[u^\mu(\partial^\nu\phi)-u^\nu(\partial^\mu\phi)\right](u\cdot\partial\phi)+\lambda_2\left[v^\mu(\partial^\nu\phi)-v^\nu(\partial^\mu\phi)\right](v\cdot\partial\phi) \  . 
\end{eqnarray}


\section{Investigating Lorentz symmetry violation in the Casimir effect}
\label{Casimir_LV}

\begin{figure}[!h]
\centering
\includegraphics[scale=0.3]{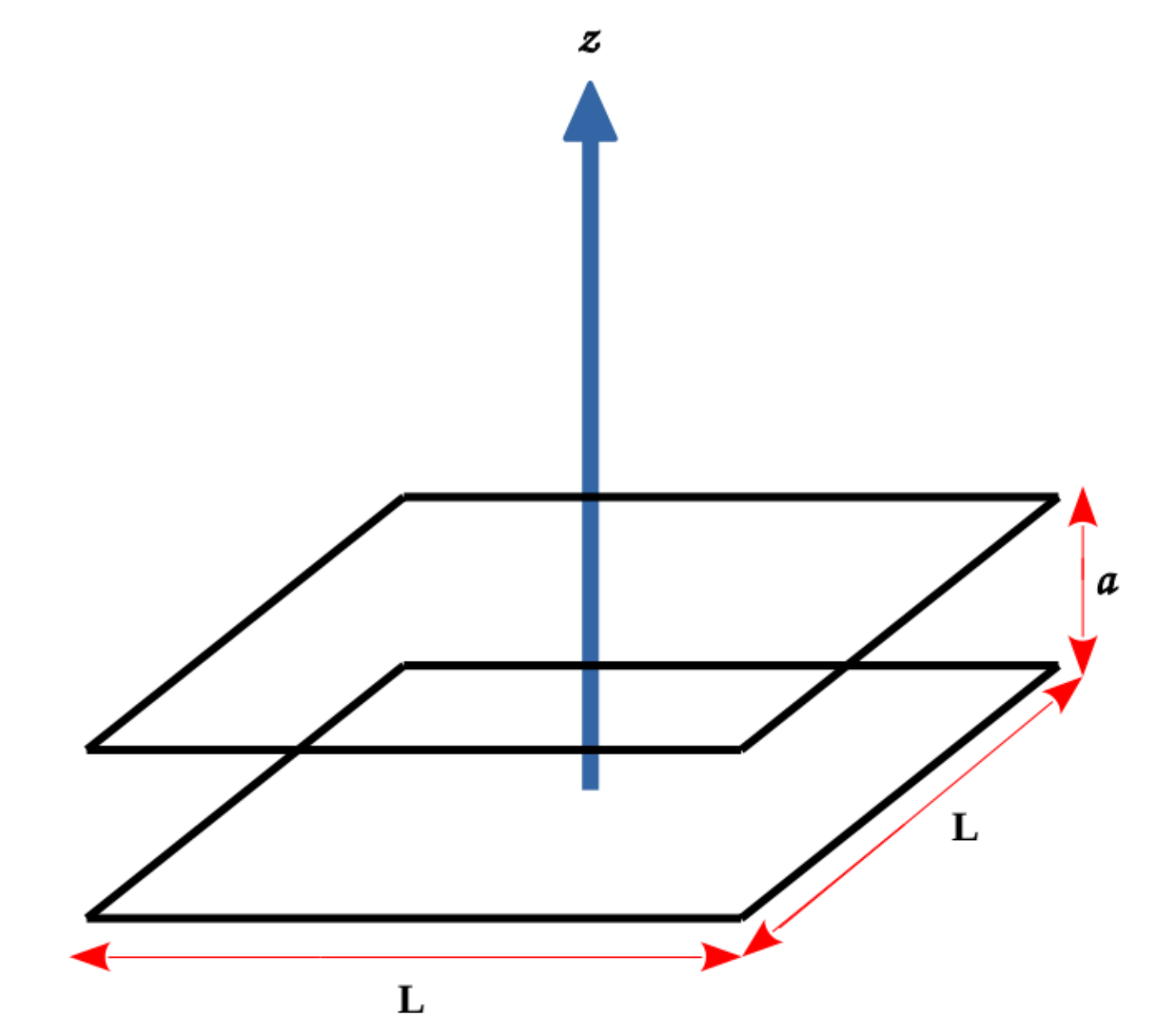}
\caption{Two parallel plates with an area of $L^{2}$ are separated by a distance $a \ll L$.}
\label{Plates}
\end{figure}

The main objective of this section is to investigate how the LV symmetry, represented by the presence of two orthogonal constant vectors in the equation governing the dynamics of the scalar quantum field, Eq. \eqref{KG_equation}, modifies the dispersion relation that is responsible for the deviation of the Casimir energy and pressure when compared with the scenario without spacetime anisotropy. As we have already mentioned, we will consider different boundary conditions imposed on the fields $\phi(x)$ on the plates, as represented in Fig. \ref{Plates}.

As previously explained, in this analysis, we assume the presence of two orthogonal constant four-vectors, $u^\mu$ and $v^\nu$, coupled to the scalar quantum field. Throughout our development, we consider $u^\mu$ as a time-like unitary four-vector, $u^\mu=(1, 0, 0 , 0)$. As for the unitary four-vector $v^\mu$, it is space-like, and we will investigate two distinct configurations: the vector $v^\mu$ parallel to the plates and perpendicular to them. Therefore, without loss of generality, we consider $v^\mu=(0, 1, 0 , 0)$ and $v^\mu=(0, 0, 0, 1)$ for the parallel and perpendicular configurations, respectively. 

In order to develop our formalism, we first need to obtain the normalized solutions of Eq. \eqref{KG_equation} satisfying specific boundary conditions on the plates. This will allow us to obtain the corresponding hamiltonian operator. Once we have this operator, we can calculate the total vacuum energy of the system and then determine the Casimir energy for each case.


\subsection{Dirichlet boundary condition}
\label{Dirichlet}

It was shown in \cite{Messias2017} that the quantum field operator $\hat{\phi}(x)$ can be expressed in terms of the normalized positive/negative energy solutions of the Klein-Gordon equation that satisfy the Dirichlet boundary condition on the plates at $z=0$ and $z=a$:
\begin{eqnarray}
 \left. \phi \left(x\right)\right|_{z=0}=\left. \phi \left(x\right)\right|_{z=a}  \  ,
\end{eqnarray}
is expressed by
\begin{eqnarray}
	\label{Sol_D}
 \hat{\phi}(x)=\int{d^{2}\vec{k}}\sum_{n=1}^{\infty}\left[\frac{1}{\left(2\pi\right)^{2}\omega_{\vec{k},n}a}\right]
 ^{\frac{1}{2}}\sin\left(\frac{n\pi}{a}z\right)\left[\hat{a}_{n}(\vec{k})e^{-ikx}+\hat{a}_{n}^{\dagger}(\vec{k})
 e^{ikx}\right]
\end{eqnarray}
where 
\begin{eqnarray}
kx=\omega_{\vec{k},n}t-k_{x}x-k_{y}y=\omega_{\vec{k},n}t-\vec{k}_{\parallel}\cdot{\vec{x}}_{\parallel}  \  .
\end{eqnarray}

In Eq. \eqref{Sol_D}, $\hat{a}_{n}(\vec{k})$ and $\hat{a}_{n}^{\dagger}(\vec{k})$ are the annihilation and creation operators, respectively, characterized by the complete set of quantum numbers $\sigma=\left \{k_{x},k_{y},n\right \}$. These operators satisfy the algebraic relations
\begin{eqnarray}
\label{vert}
 \left[\hat{a}_{n}(\vec{k}),\hat{a}_{n'}^{\dagger}(\vec{k'})\right]&=&\delta_{n,n'}\delta^{2}(\vec{k}_{\parallel}-\vec{k'}_{\parallel})  \ \ \ \text{and} \ \ \ 
 \left[\hat{a}_{n}(\vec{k}),\hat{a}_{n'}(\vec{k'})\right] = \left[\hat{a}_{n}^{\dagger}(\vec{k}),\hat{a}_{n'}^{\dagger}(\vec{k'})\right]=0. 
\end{eqnarray}

\subsubsection{First model}

In this analysis, we will consider $u^\mu=(1, 0, 0, 0 )$ and $v^{\mu}=(0, 1 , 0, 0)$. For this case, the general expression \eqref{Sol_D} will be a solution of the Klein-Gordon equation, Eq. \eqref{KG_equation}, since it satisfies the following dispersion relation:
\begin{eqnarray}
	\label{Disp_1}
 \omega^{2}_{\vec{k},n}=\frac{1}{(1+\lambda_1)}\left[(1-\lambda_2)k_{x}^{2}+k_{y}^{2}+\left(\frac{n\pi}{a}\right)^2+\mu^2\right].
\end{eqnarray}

The hamiltonian operator, $\hat{H}$, which is given by the integral of $T^{00}$, takes the form:
\begin{eqnarray}
\begin{aligned}
 \hat{H}=&\frac{1}{2}\int{d^{3}\vec{x}}\left[(1+\lambda_1)(\partial_{t}\hat{\phi})^{2}+(\vec{\nabla}\hat{\phi})^{2}-\lambda_2(\partial_x{\hat\phi})^2+\mu^2\hat{\phi}^2\right], \\
 =&
 \frac{(1+\lambda_1)}{2}\int{d^{2}\vec{k}}\sum_{n=1}^{\infty}\omega_{\vec{k},n}\left[2\hat{a}_{n}^{\dagger}(\vec{k})
 \hat{a}_{n}(\vec{k})+\frac{L^2}{(2\pi)^2}\right].
\end{aligned}
\end{eqnarray}
Consequently, the vacuum energy is obtained by taking the vacuum expectation value of $\hat{H}$:
\begin{eqnarray}
\label{EVDt}
 E_{0}=\left<\right.0|\hat{H}|\left.0\right>=\frac{(1+\lambda_1)L^{2}}{8\pi^{2}}\int{d^{2}\vec{k}}\sum_{n=1}^{\infty}
 \omega_{\vec{k},n} .
\end{eqnarray}

To perform the summation over the quantum number $n$ in the above expression, we will use the Abel-Plana formula below \cite{Bordag:2009zzd}: 
\begin{eqnarray}
\label{AP1}
\sum_{n=0}^{\infty}F(n)=\frac{1}{2}F(0)+\int_{0}^{\infty}{du}F(u)+i\int_{0}^{\infty}\frac{du}{e^{2\pi u}-1}\left[F(iu)-
F(-iu)\right],
\end{eqnarray}
where for our purpose, we have
\begin{eqnarray}
	F(n)=\frac1{\sqrt{1+\lambda_1}}\left[(1-\lambda_2)k_x^2+k_y^2+\left(\frac{n\pi}{a}\right)^2+\mu^2\right]^{1/2} .
\end{eqnarray}
To develop the integral over the plane $(k_{x}, k_{y})$, we redefine $\bar{k}_x=k_x\sqrt{1-\lambda_2}$ and change $d\bar{k}_x,dk_y$ to polar coordinates. The first contribution obtained by substituting Eq. \eqref{AP1} into Eq. \eqref{EVDt} provides the vacuum energy in the presence of a single plate, while the second corresponds to the vacuum energy without boundaries; these results are divergent and do not contribute to the Casimir energy, so they will be discarded in our analysis. The Casimir energy will be given by the third contribution. Finally, defining a new variable of integration $w=\pi u/a$, we obtain:
\begin{eqnarray}
\begin{aligned}
\frac{E_C}{L^2} = i\frac{\sqrt{1+\lambda_1}}{\sqrt{1-\lambda_2}}\frac a{4\pi^2}\int_0^\infty dk k \int_0^\infty \frac{dw}{e^{2aw}-1}  & \left[\sqrt{k^2+(iw)^2+\mu^2} \right. \\ & \ \left. - \sqrt{k^2+(-iw)^2+\mu^2}\right].
\end{aligned}
\end{eqnarray} 

The integral over $w$ must be considered in two different segments: in the first one, $[0, \sqrt{k^2+\mu^2}]$, the integral vanishes, so only the contribution coming from the second segment, $[\sqrt{k^2+\mu^2}, \infty)$, remains. Taking this into account, we obtain:
\begin{eqnarray}
 \frac{E_{C}}{L^{2}}=-\sqrt{\frac{1+\lambda_1}{1-\lambda_2}}\frac a{2\pi^2} \int_{0}^{\infty}{kdk}\int_{\sqrt{k^2+\mu^2}}^{\infty}
 \frac{dw}{e^{2aw}-1}\sqrt{w^2-(k^2+\mu^2)}.
\end{eqnarray}
Performing another change of variable, $\rho^{2}=w^{2}-(k^{2}+\mu^{2})$, we obtain:
\begin{eqnarray}
 \frac{E_{C}}{L^{2}}=-\sqrt{\frac{1+\lambda_1}{1-\lambda_2}}\frac a{2\pi^2} \int_{0}^{\infty}{kdk}\int_{0}^{\infty}d\rho{\frac{\rho^{2}}{
 \sqrt{\rho^{2}+k^{2}+\mu^{2}}\left(e^{2a\sqrt{\rho^{2}+k^{2}+\mu^{2}}}-1\right)}}.
\end{eqnarray}
Finally, by changing the coordinates $(k, \rho)$ to polar coordinates $(\sigma, \phi)$ and then evaluating the integral over the angular variable, we obtain:
\begin{eqnarray}
	\label{EC_1a}
 \frac{E_{C}}{L^{2}}=-\sqrt{\frac{1+\lambda_1}{1-\lambda_2}}\frac a{6\pi^2} \int_{0}^{\infty}d\sigma\frac{\sigma^{4}}{\sqrt{\sigma^{2}+\mu^{2}}
 \left(e^{2a\sqrt{\sigma^{2}+\mu^{2}}}-1\right)}.
\end{eqnarray}
As we can see, when $\lambda_2=-\lambda_1$, there is a cancellation in the correction to the Casimir energy caused by LV. However, this situation is very unlikely, and we can approximate the correction to the Casimir energy caused by the presence of these two vectors to be given by the overall factor $\sqrt{1+(\lambda_1+\lambda_2)}$.

Unfortunately, the above integral cannot be expressed in terms of elementary functions for $\mu\neq 0$; however, in the massless field limit, it can be expressed as follows:
\begin{eqnarray}
	\frac{E_{C}}{L^{2}}=-\sqrt{\frac{1+\lambda_1}{1-\lambda_2}}\frac{\pi^2}{1440a^3}.
\end{eqnarray}
In Eq. \eqref{EC_1a}, we can perform another change of variable, $\xi^2=\sigma^2+\mu^2$, and after that, by defining $\xi=\mu v$, we get:
\begin{eqnarray}
\label{int1}
 \frac{E_{C}}{L^{2}}=-\sqrt{\frac{1+\lambda_1}{1-\lambda_2}}\frac {a\mu^4}{6\pi^2} \int_{1}^{\infty}\frac{(v^{2}-1)^{\frac{3}{2}}dv}
 {e^{2a\mu v}-1}  \  .
\end{eqnarray}
From the above result, we can observe that the correction to the Casimir energy caused by the LV is given by a multiplicative factor.

Using the expansion $(e^{x}-1)^{-1}=\sum_{l=1}^\infty e^{-lx}$, we can express \eqref{int1} in terms of an infinite sum of Bessel functions \cite{Abra}:
\begin{eqnarray}
	\label{int2}
	\frac{E_{C}}{L^{2}}=-\sqrt{\frac{1+\lambda_1}{1-\lambda_2}}\frac {\mu^2}{8\pi^2a}\sum_{l=1}^\infty \frac{K_2(2la\mu)}{l^2}   \  .
\end{eqnarray}
\begin{itemize}
\item For $a\mu \gg 1$, the dominant term in the above expression is the component with $l=1$. By using the asymptotic expression for the Bessel function, we obtain:
\begin{eqnarray}
	\label{int3}
	\frac{E_{C}}{L^{2}}\approx-\sqrt{\frac{1+\lambda_1}{1-\lambda_2}}\frac 1{16}\left(\frac\mu{\pi a}\right)^{3/2}e^{-2a\mu},
\end{eqnarray}
the Casimir energy decays exponentially with $\mu a$.

\item For $a\mu \ll 1$, we can approximate the integrand in Eq. \eqref{int1} as shown below and obtain a series expansion:
\begin{eqnarray}
	\label{appr_1}
	\frac{E_C}{L^2}&\approx& -\sqrt{\frac{1+\lambda_1}{1-\lambda_2}}\frac {a\mu^4}{6\pi^2}\int_1^\infty dv\frac{(v^3-\frac{3v}{2})} {e^{2a\mu v}-1} \nonumber\\
	&\approx&-\sqrt{\frac{1+\lambda_1}{1-\lambda_2}}\frac1{1440\pi^2}\frac{1}{a^3}\left[\pi^4-15\pi^2(a\mu)^2+140(a\mu)^3\right].
\end{eqnarray}
From the above expression, the Casimir pressure between the two parallel plates due to the scalar field takes the form:
\begin{eqnarray}
P_C(a)\approx-\sqrt {{\frac {1+\lambda_{{1}}}{1-\lambda_{{2}}}}}\frac{1}{480}{\frac {\left[\pi^2- 5(\mu a)^{2} \right] }{{a}^{4}}}.
\end{eqnarray}
\end{itemize}

\subsubsection{Second model}
\label{Perpendicular}

In this subsection, we will keep the same vector $u^\mu$ but consider $v^\mu$ orthogonal to the plates:
\begin{eqnarray}
 v^{\mu}=(0, 0, 0, 1).
\end{eqnarray}
For this case, the dispersion relation is obtained by direct substitution of the field operator Eq. \eqref{Sol_D} into Eq. \eqref{KG_equation}. It is given by:
\begin{eqnarray}
	\label{Disp_2}
 \omega^{2}_{\vec{k},n}=\frac1{1+\lambda_1}\left[k_{x}^{2}+k_{y}^{2}+(1-\lambda_2)\left(\frac{n\pi}{a}\right)^2+\mu^2\right].
\end{eqnarray}

The hamiltonian operator is now given by:
\begin{eqnarray}
 \hat{H}=\frac{1}{2}\int{d^{3}\vec{x}}\left[(1+\lambda_1)(\partial_{t}\hat{\phi})^{2}+(\vec{\nabla}\hat{\phi})^{2}-\lambda_2(
 \partial_{z}\hat{\phi})^{2}+\mu^2\hat{\phi}^2\right] \  .
 \end{eqnarray}
Once again, substituting the field operator into the above expression and using the dispersion relation Eq. \eqref{Disp_2}, we obtain:
\begin{eqnarray}
\hat{H}=\frac{(1+\lambda_1)}{2}\int{d^{2}\vec{k}}\sum_{n=1}^{\infty}\omega_{\vec{k},n}\left[2\hat{a}_{n}^{\dagger}(\vec{k})
 \hat{a}_{n}(\vec{k})+\frac{L^2}{(2\pi)^2}\right],
\end{eqnarray}
consequently, gives the vacuum energy:
\begin{eqnarray}
 E_{0}=\left<\right.0|\hat{H}|\left.0\right>=\frac{(1+\lambda_1)L^{2}}{8\pi^{2}}\int{d^{2}\vec{k}}\sum_{n=1}^{\infty}
 \omega_{\vec{k},n}.
\end{eqnarray}
In order to again perform the summation over the integer quantum number $n$ using Eq. \eqref{AP1}, we take
\begin{eqnarray}
	F(n)=\frac1{\sqrt{1+\lambda_1}}\left[k^2+\left(\frac{n\pi}{b}\right)^2+\mu^2\right]^{\frac{1}{2}},
\end{eqnarray}
where
\begin{eqnarray}
	\label{b-factor}
	b=\frac{a}{\sqrt{1-\lambda_2}}.
\end{eqnarray}

Performing a change of coordinates from $(k_x, k_y)$ to polar coordinates and using a similar procedure as in the previous case, we find that the Casimir energy can be evaluated as follows:
\begin{eqnarray}
 \frac{E_{C}}{L^{2}}=\frac{\sqrt{1+\lambda_1}}{4\pi}i\int_{0}^{\infty}{kdk}\int_{0}^{\infty}dt
\frac{\left[k^2+\mu^2+\left(\frac{it\pi}{b}\right)^2\right]^{\frac{1}{2}}-\left[k^2+\mu^2+\left(\frac{-it\pi}{b}\right)^2
	\right]^{\frac{1}{2}}}{e^{2\pi t}-1} .
\end{eqnarray}
Following a similar procedure to the one adopted in our last subsection, we arrive at
\begin{eqnarray}
	\label{int3a}
	\frac{E_{C}}{L^{2}}=-\frac{\sqrt{1+\lambda_1}b\mu^4}{6\pi^{2}}\int_{1}^{\infty}\frac{(v^{2}-1)^{\frac{3}{2}}dv}
	{e^{2b\mu v}-1}.
\end{eqnarray}
As we can see, in this case the modification of the Casimir energy due to the LV is more delicate. It appears as a multiplicative factor through the term $\sqrt{1+\lambda_1}$, but it is also present in the integrand through the parameter $b=a/\sqrt{1-\lambda_2}$. Using the expansion $(e^{x}-1)^{-1}=\sum_{l=1}^\infty e^{-lx}$, we can express \eqref{int3a} in terms of an infinite sum of modified Bessel functions,
\begin{eqnarray}
	\label{int4a}
	\frac{E_{C}}{L^{2}}=-\frac {\sqrt{1+\lambda_1}\mu^2}{8\pi^2b}\sum_{l=1}^\infty \frac{K_2(2lb\mu)}{l^2}   \  .
\end{eqnarray}

\begin{figure}[h!]
    \centering
    	\subfigure{\includegraphics[width=0.4\textwidth]{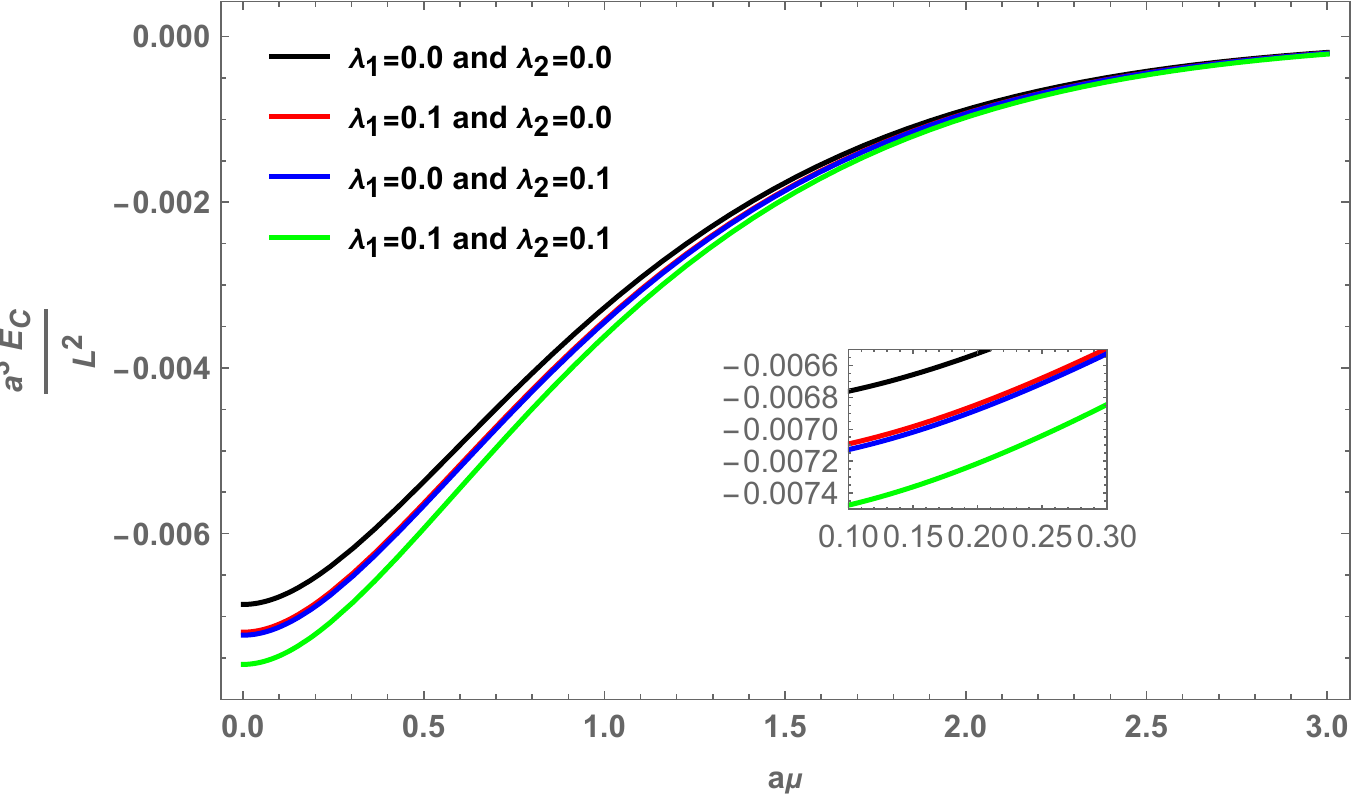}} 
    \subfigure{\includegraphics[width=0.4\textwidth]{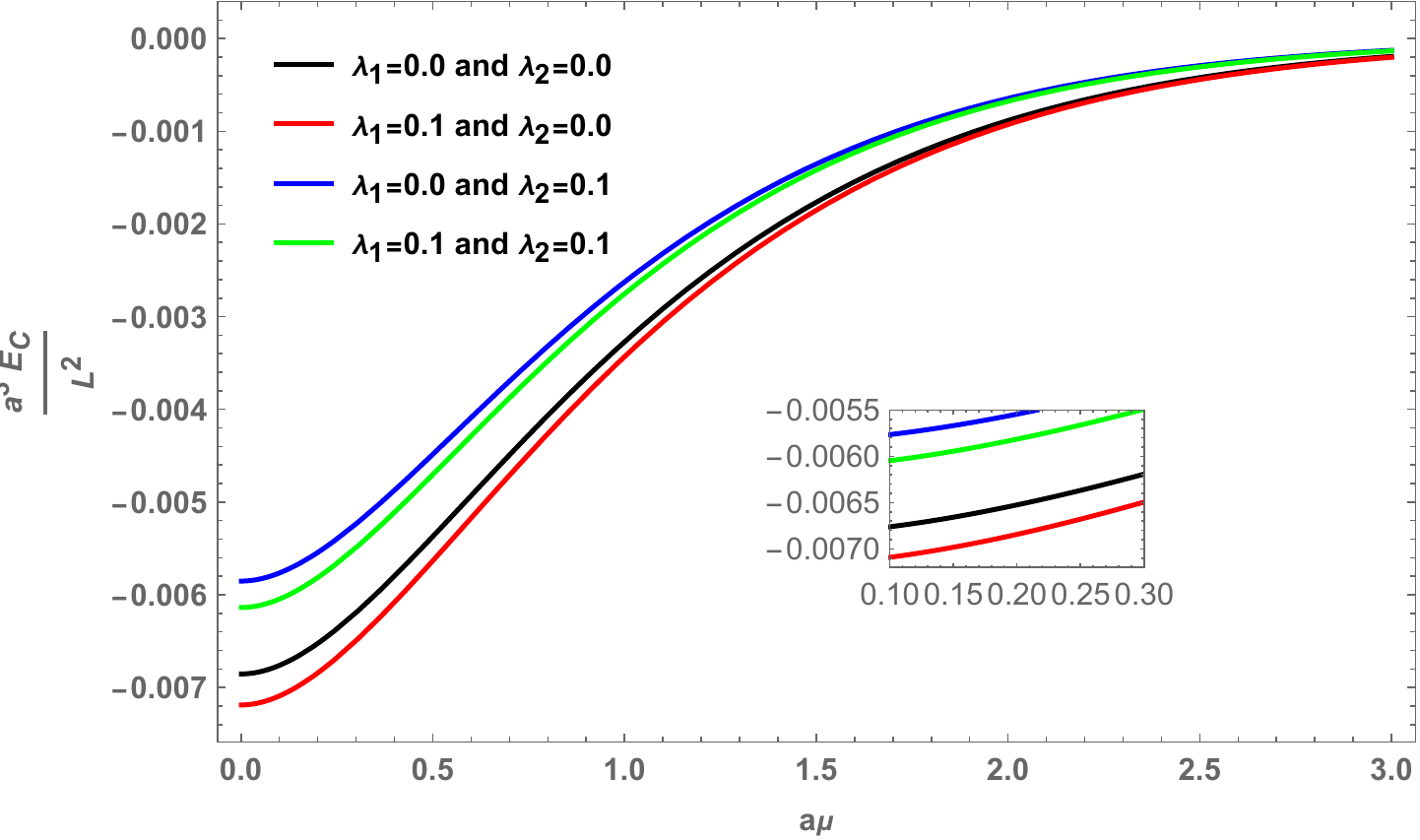}} 
    \caption{The behaviors of the Casimir energies per unit area multiplied by $a^3$ as a function of $a\mu$, adopting different values of the parameters $\lambda_1$ and $\lambda_2$, are presented in the graph. In the left panel, we consider the four-vector $v^\mu$ parallel to the plates, and in the right panel, we consider this vector perpendicular to them.}
    \label{plot_dirichlet}
\end{figure}

We can obtain approximate expressions for the Casimir energy in the limits: $b\mu \gg 1$ and $b\mu \ll 1$. 
\begin{itemize}
\item For the first case, we can take the result Eq. \eqref{int4a} and use the asymptotic expression for the modified Bessel function for large arguments \cite{Abra}. As before, the dominant contribution comes from the term $l=1$. The result is:
\begin{eqnarray}
		\label{int4b}
	\frac{E_{C}}{L^{2}}\approx-\frac {\sqrt{1+\lambda_1}}{16}\left(\frac\mu{\pi b}\right)^{3/2}e^{-2b\mu}   \  .
\end{eqnarray}

\item For $b\mu\ll1$, we can approximate the integrand in Eq. \eqref{int3a} as shown in the last subsection. Taking into account Eq. \eqref{b-factor}, the result is: 
\begin{eqnarray}
	\label{int4c}
	\frac{E_C}{L^2}\approx-\frac{\sqrt{1+\lambda_1}}{1440}\frac{(1-\lambda_2)^{3/2}}{a^3\pi^2}\left[\pi^4-\frac{15\pi^2(a\mu)^2}{1-\lambda_2}+ \frac{140(a\mu)^3}{(1-\lambda_2)^{3/2}}\right]  \  .
\end{eqnarray}
As for the Casimir pressure, we obtain,
\begin{eqnarray}
	P_C(a)\approx-\frac{\sqrt{1+\lambda_1}}{480}\frac{(1-\lambda_2)^{3/2}}{a^4}\left[\pi^2- \frac{5(\mu a)^{2}}{1-\lambda_2} \right]  \  .
\end{eqnarray}
\end{itemize}
In Fig. \ref{plot_dirichlet}, we show the behaviors of the Casimir energies as a function of $a\mu$, considering different values for the parameters $\lambda_1$ and $\lambda_2$, for the two models analyzed. In the left panel, we present the behavior associated with $v^\mu$ parallel to the plates, and in the right panel, we show the behavior for the vector $v^\mu$ perpendicular to them.

Although we have mentioned at the beginning of this paper that we will consider only one time-like vector, $u^\mu$, and only one space-like vector $v^\mu$, the result found above, Eq. \eqref{int3a} can be extended for the case where the vector $u^\mu$ is also space-like. Considering this vector parallel to the plates, the result obtained fore the LV Casimir energy, is given by,
\begin{eqnarray}
	\label{int3a1}
	\frac{E_{C}}{L^{2}}=-\frac{b\mu^4}{6\pi^{2}\sqrt{1-\lambda_1}}\int_{1}^{\infty}\frac{(v^{2}-1)^{\frac{3}{2}}dv}
	{e^{2b\mu v}-1}  \  .
\end{eqnarray}
Because the parameter $\lambda_1$ is smaller thatn unity, the first order term in the expansion  $\frac1{\sqrt{1-\lambda_1}}$ in powers of $\lambda_1$, would provide the same correction on the Casimir energy as given in  \eqref{int3a}.

\subsection{Neumann boundary condition}
\label{Neumann}

The general expression for the quantum field operator $\hat{\phi}$ that satisfies the Neumann boundary condition on the two plates is:
\begin{eqnarray}
	\left.\frac{\partial \phi(x)}{\partial z}\right|_{z=0}=\left.\frac{\partial \phi(x)}{\partial z}\right|_{z=a}=0 ,
\end{eqnarray}
was given in \cite{Messias2017}. For this case, the scalar field operator reads,
\begin{eqnarray}
	\hat{\phi}(x)=\int{d^{2}\vec{k}}\sum_{n=0}^{\infty}c_{n}\cos \left(\frac{n\pi}{a}z\right)\left[\hat{a}_{n}(\vec{k})
	e^{-ikx}+\hat{a}_{n}^{\dagger}(\vec{k})e^{ikx}\right] ,
\end{eqnarray}
where the normalization constant is 
\begin{eqnarray}
	\label{vert}
	c_{n}=\left \{\begin{array}{c}
		\left[\frac{1}{2(2\pi)^{\frac{1}{2}}\omega_{\vec{k},n}a}\right]^{\frac{1}{2}} \text{ }\text{for} \text{ }n=0, \\ \\
		\left[\frac{1}{(2\pi)^{\frac{1}{2}}\omega_{\vec{k},n}a}\right]^{\frac{1}{2}} \text{ } \text{for} \text{ } n\geq 0.
	\end{array} \right.
\end{eqnarray}

Although the structure of the field operator here is different from the corresponding one for the Dirichlet case, given by Eq. \eqref{Sol_D}, the Hamiltonian operator and the dispersion relations are the same for both models considered in section \ref{Dirichlet}. Therefore, we decided not to present all the calculations in this section, since they provide exactly the same results for the Casimir energies and pressures as in the previous section.


\subsection{Mixed boundary conditions}
\label{Mixed}

Now, let us consider the case where the scalar quantum field obeys Dirichlet boundary condition on one plate and Neumann on the other. In this case, we have two different configurations for the scalar quantum field:

\begin{itemize}
 \item First configuration:
 \begin{eqnarray}
  \left.\phi(x)\right|_{z=0}=\left.\frac{\partial \phi(x)}{\partial z}\right|_{z=a}=0.
 \end{eqnarray}
 \item Second configuration:
 \begin{eqnarray}
  \left.\frac{\partial \phi(x)}{\partial z}\right|_{z=0}=\left.\phi(x)\right|_{z=a}=0.
 \end{eqnarray}
\end{itemize}

The solutions of the Klein-Gordon equation, Eq. \eqref{KG_equation}, that are compatible with these conditions were obtained in \cite{Messias2017}. They are given by:
\begin{eqnarray}
\hat{\phi}_{1}(x)=\int{d^{2}\vec{k}}\sum_{n=0}^{\infty}\left[\frac{1}{(2\pi)^{2}\omega_{\vec{k},n}a}\right]^{\frac{1}{2}}
\sin \left[\left(n+\frac{1}{2}\right)\frac{\pi}{a}z\right]\left[\hat{a}_{n}(\vec{k})e^{-ikx}+\hat{a}^{\dagger}_{n}(\vec{k})
e^{ikx} \right]
\end{eqnarray}
for the first configuration, and for the second configuration, we have:
\begin{eqnarray}
\hat{\phi}_{2}(x)=\int{d^{2}\vec{k}}\sum_{n=0}^{\infty}\left[\frac{1}{(2\pi)^{2}\omega_{\vec{k},n}a}\right]^{\frac{1}{2}}
\cos \left[\left(n+\frac{1}{2}\right)\frac{\pi}{a}z\right]\left[\hat{a}_{n}(\vec{k})e^{-ikx}+\hat{a}^{\dagger}_{n}(\vec{k})
e^{ikx} \right] .
\end{eqnarray}
However, the field operators $\hat{\phi}_1$ and $\hat{\phi}_2$ give rise to the same Hamiltonian operator and dispersion relations.

\subsubsection{First model}

Let us start by considering that the four-vectors are given by $u^{\mu} = (1, 0, 0, 0)$ and $v^{\mu} = (0,1,0,0)$. For both configurations of the field operator, the hamiltonian operator reads:
\begin{eqnarray}
\begin{aligned}
 \hat{H} = \frac{(1+\lambda_1)}{2}\int{d^{2}\vec{k}}\sum_{n=0}^{\infty}\omega_{\vec{k},n}\left[2\hat{a}_{n}^{\dagger}(\vec{k}) \hat{a}_{n}(\vec{k})+\frac{L^2}{(2\pi)^2}\right] .
\end{aligned}
\end{eqnarray}
As for the dispersion relation, $\omega_{\vec{k},n}$, we obtain
\begin{eqnarray}
 \omega^{2}_{\vec{k},n}=\frac{1}{(1+\lambda_1)}\left[(1-\lambda_2)k_{x}^{2}+k_{y}^{2}+\left(\left(n+\frac{1}{2} \right)\frac{\pi}{a}\right)^2+\mu^2\right].
\end{eqnarray}
Therefore, the vacuum energy is given by:
\begin{eqnarray}
 E_{0}=\left<\right.0|\hat{H}|\left.0\right>=\frac{(1+\lambda_1)L^{2}}{8\pi^{2}}\int{d^{2}\vec{k}}\sum_{n=0}^{\infty}
 \omega_{\vec{k},n}.
\end{eqnarray}

Again, we need to perform the summation over the quantum number $n$, but for this case, we can use the Abel-Plana formula for half-integer numbers, given below \cite{Bordag:2009zzd}:
\begin{eqnarray}
\begin{aligned} \label{AP_2}
    \sum_{n=0}^{\infty}F\left(n+\frac{1}{2}\right) = \int_{0}^{\infty}F(u)du - i\int_{0}^{\infty}\frac{du}{e^{2\pi u}+1}\left[F(iu)-F(-iu)\right] ,
\end{aligned}
\end{eqnarray}
where
\begin{eqnarray}
\begin{aligned}
    F\left(n+\frac{1}{2}\right) = \frac{1}{\sqrt{1+\lambda_1}}\left[(1-\lambda_2)k_{x}^{2}+k_{y}^{2}+\left(\left(n+{1}/{2} \right)\frac{\pi}{a}\right)^2+\mu^2\right]^{1/2} .
\end{aligned}
\end{eqnarray}
As we have already mentioned, the first term in Eq. \eqref{AP_2} is discarded because it corresponds to the free vacuum energy. Thus, the Casimir energy is given by the second term. Defining $\bar{k_x}=k_x\sqrt{1-\lambda_2}$ and changing $d{\bar{k_x}}dk_y$ to polar coordinates, we obtain:
\begin{eqnarray}
\begin{aligned}
    \frac{E_C}{L^2}= - i \frac{\sqrt{1+\lambda_1}}{\sqrt{1-\lambda_2}}\frac a{4\pi^2}\int_0^\infty dk k \int_0^\infty \frac{dw}{e^{2aw}+1} & \left[\sqrt{k^2+(iw)^2+\mu^2} \right. \\ & \left. -\sqrt{k^2+(-iw)^2+\mu^2}\right] .
\end{aligned}
\end{eqnarray}
In the above equation, we changed the integration variable to $w=\pi u / a$. To perform the integral over $w$, we consider two different segments: the first integration interval is $[0, \sqrt{k^2 + \mu^2}]$, where the integral vanishes. Therefore, the contribution comes only from the second segment, $[\sqrt{k^2 + \mu^2}, \infty)$. Thus, we obtain:
\begin{eqnarray}
\begin{aligned}
    \frac{E_{C}}{L^{2}} = \sqrt{\frac{1+\lambda_1}{1-\lambda_2}}\frac a{2\pi^2} \int_{0}^{\infty}{kdk}\int_{\sqrt{k^2+\mu^2}}^{\infty}
    \frac{dw}{e^{2aw}+1}\sqrt{w^2-(k^2+\mu^2)} .
\end{aligned}
\end{eqnarray}
Performing another change of variable, $\rho^2 = w^2-k^2-\mu^2$, we obtain:
\begin{eqnarray}
\begin{aligned}
    \frac{E_{C}}{L^{2}} = \sqrt{\frac{1+\lambda_1}{1-\lambda_2}}\frac a{2\pi^2} \int_{0}^{\infty}{kdk}\int_{0}^{\infty}
    \frac{\rho^2 d\rho}{\sqrt{\rho^2+k^2+\mu^2} \left(e^{2a \sqrt{\rho^2+k^2+\mu^2}}+1 \right)} .
\end{aligned}
\end{eqnarray}

At this point, we make a change of coordinates from $(k, \rho)$ to polar coordinates $(\sigma, \phi)$ and then define a new variable $\xi^2 = \sigma^2 + \mu^2$ and finally $\xi = \mu v$. The result is:
\begin{eqnarray} \label{casimir_energy_exact_mixed_1}
\begin{aligned}
    \frac{E_{C}}{L^{2}} = \sqrt{\frac{1+\lambda_1}{1-\lambda_2}} \frac{a\mu^4}{6\pi^2} \int_{1}^{\infty} \frac{(v^2-1)^{\frac{3}{2}}}{e^{2a\mu v}+1} dv .
\end{aligned}
\end{eqnarray}
Using the series expansion $(e^x+1)^{-1}=\sum_{l=1}^\infty(-1)^{l-1} e^{-lx}$, we can express the above integral in terms of modified Bessel functions:
\begin{eqnarray}
	\label{casimir_energy_exact_mixed_1a}
\frac{E_{C}}{L^{2}} =- \sqrt{\frac{1+\lambda_1}{1-\lambda_2}} \frac{\mu^2}{8\pi^2 a}\sum_{l=1}^\infty(-1)^l\frac{K_2(2la\mu)}{l^2} .	
\end{eqnarray}
Here we can also obtain approximate expressions for the Casimir energy in the limits $a\mu \gg 1$ and $a\mu \ll 1$.
\begin{itemize}
\item For the first limit, the dominant contribution in Eq. \eqref{casimir_energy_exact_mixed_1a} comes from the term $l=1$. It is given by:
\begin{eqnarray}
	\label{Mixed_1}
		\frac{E_{C}}{L^{2}} \approx \sqrt{\frac{1+\lambda_1}{1-\lambda_2}} \frac{1}{16} \left(\frac{\mu}{\pi a}\right)^{\frac{3}{2}} e^{-2 a \mu} ,
\end{eqnarray}
As we can see, in this case, the Casimir energy tends to zero very quickly.
\item For the second limit, $a \mu \ll 1$, we use the approximate expression for Eq. \eqref{casimir_energy_exact_mixed_1}:
\begin{eqnarray}
\label{Mixed_2}
\begin{aligned}
    \frac{E_{C}}{L^{2}} & \approx \sqrt{\frac{1+\lambda_1}{1-\lambda_2}} \frac{a\mu^4}{6\pi^2} \int_{1}^{\infty} \frac{v^3-\frac{3 v}{2}}{e^{2a\mu v}+1} dv   \\
    & \approx \sqrt{\frac{1+\lambda_1}{1-\lambda_2}} \frac{1}{11520 a^3} \left(7\pi^2- 60 a^2 \mu^2\right) .
\end{aligned}
\end{eqnarray}
As a result, the Casimir pressure between the two parallel plates takes the form:
\begin{eqnarray}
\begin{aligned}
    P_{C} = \sqrt{\frac{1+\lambda_1}{1-\lambda_2}} \frac{1}{3840  a^4} \left(7\pi^2 - 20  a^2 \mu^2  \right) .
\end{aligned}
\end{eqnarray}
\end{itemize}

\subsubsection{Second model}

As mentioned before, we will continue with the same four-vector $u^\mu$; however, for the second one, we take it to be orthogonal to the plates:
\begin{eqnarray}
\begin{aligned}
    v^\mu=(0, 0, 0, 1) .
\end{aligned}
\end{eqnarray}
Thus, the hamiltonian operator is given by:
\begin{eqnarray} \label{hamiltonian_mixed_2}
\begin{aligned}
    \hat{H} = \frac{(1+\lambda_1)}{2}\int{d^{2}\vec{k}}\sum_{n=0}^{\infty}\omega_{\vec{k},n}\left[2\hat{a}_{n}^{\dagger}(\vec{k}) \hat{a}_{n}(\vec{k})+\frac{L^2}{(2\pi)^2}\right] ,
\end{aligned}
\end{eqnarray}
where the dispersion relation is given by
\begin{eqnarray} \label{dispersion_mixed_2}
\begin{aligned}
    \omega^{2}_{\vec{k},n}=\frac{1}{(1+\lambda_1)}\left[k_{x}^{2}+k_{y}^{2}+(1-\lambda_2)\left[\left(n+\frac{1}{2} \right)\frac{\pi}{a}\right]^2+\mu^2\right] .
\end{aligned}
\end{eqnarray}
Therefore, from Eq. \eqref{hamiltonian_mixed_2}, the vacuum energy is given by:
\begin{eqnarray}
\begin{aligned}
    E_{0} = \left<\right.0|\hat{H}|\left.0\right>=\frac{(1+\lambda_1)L^{2}}{8\pi^{2}}\int{d^{2}\vec{k}}\sum_{n=0}^{\infty}
    \omega_{\vec{k},n} .
\end{aligned}
\end{eqnarray}

Here, we also use the Abel-Plana formula given in Eq. \eqref{AP_2}, together with the dispersion relation, Eq. \eqref{dispersion_mixed_2}, to evaluate the sum over $n$. For this purpose, we have,
\begin{eqnarray}
\begin{aligned}
    F\left(n+\frac{1}{2}\right) = \frac{1}{\sqrt{1+\lambda_1}}\left[k_{x}^{2}+k_{y}^{2}+\left(\left(n+\frac{1}{2} \right)\frac{\pi}{b}\right)^2+\mu^2\right]^{1/2} 
\end{aligned}
\end{eqnarray}
where
\begin{eqnarray} \label{parameter_b}
    b = \frac{a}{\sqrt{1-\lambda_2}} .
\end{eqnarray}
Again, after proceeding with the renormalization and discarding the first term of the sum in the Abel-Plana formula, the Casimir energy per unit area is expressed as:
\begin{eqnarray}
\begin{aligned}
    \frac{E_C}{L^2}= - i \sqrt{1+\lambda_1} \frac{b}{4\pi^2}\int_0^\infty dk k \int_0^\infty \frac{dw}{e^{2bw}+1} & \left[\sqrt{k^2+(iw)^2+\mu^2} \right. \\ & \left. -\sqrt{k^2+(-iw)^2+\mu^2}\right] .
\end{aligned}
\end{eqnarray} 
In the above equation, we performed a change of variable $w=\pi u/b$ and then a transformation to polar coordinates. After some intermediate steps, the Casimir energy per unit area is expressed as:
\begin{eqnarray}
\begin{aligned}
    \frac{E_{C}}{L^{2}} = \sqrt{1+\lambda_1}\frac{b}{2\pi^2} \int_{0}^{\infty}{kdk}\int_{\sqrt{k^2+\mu^2}}^{\infty}
    \frac{dw}{e^{2bw}+1}\sqrt{w^2-(k^2+\mu^2)} .
\end{aligned}
\end{eqnarray}
And finally, following similar steps adopted in previous sections, we obtain:
\begin{eqnarray} 
\label{casmir_energy_mixed_2}
\begin{aligned}
    \frac{E_{C}}{L^{2}} = \sqrt{1+\lambda_1} \frac{b \mu^4}{6\pi^2} \int_{1}^{\infty} \frac{(v^2-1)^{\frac{3}{2}}}{e^{2 b \mu v}+1} dv .
\end{aligned}
\end{eqnarray}
We can observe once more that the modification in the Casimir energy due to the Lorentz violation is through a multiplicative factor and also in the integrand function.\footnote{A similar discussion as is in the final of subsection \ref{Perpendicular} is also applicable here.}

Using the series expansion $(e^x+1)=-\sum_{l=1}^\infty(-1)^l e^{-lx}$, we can express the above result as:
\begin{eqnarray}
\label{casimir_energy_exact_mixed_1b}
	\frac{E_{C}}{L^{2}} =- \sqrt{1+\lambda_1} \frac{\mu^2}{8\pi^2 b}\sum_{l=1}^\infty(-1)^l\frac{K_2(2lb\mu)}{l^2} .	
\end{eqnarray}
Below, we present approximate expressions for Eq. \eqref{casimir_energy_exact_mixed_1b} in the asymptotic limits:
\begin{itemize}
\item For the case $b \mu \gg 1$, we have:
\begin{eqnarray}
	\label{Mixed_3}
	\frac{E_{C}}{L^{2}}  \approx \frac{\sqrt{1+\lambda_1}}{16} \left(\frac{\mu}{\pi b}\right)^{\frac{3}{2}} e^{-2 b \mu} ,
\end{eqnarray}
i.e., in this limit the Casimir energy tends to zero very fast.
\item In the case $b \mu \ll 1$, using the approximate expression for the integrand of Eq. \eqref{casmir_energy_mixed_2}, we obtain: 
\begin{eqnarray} 
	\label{casmir_energy_mixed_2_small}
	\frac{E_{C}}{L^{2}} \approx \frac{\sqrt{1+\lambda_1}}{11520 b^3} \left[7\pi^2 - 60  (b \mu)^2\right] .
\end{eqnarray}
Note that the above result depends on the parameter $\lambda_2$ through the parameter $b$ given in Eq. \eqref{parameter_b}. Therefore, the Lorentz violation parameters in this result are more delicate. So, considering Eq. \eqref{parameter_b}, we get:
\begin{eqnarray}
\begin{aligned}
    \frac{E_{C}}{L^{2}} = \frac{\sqrt{1+\lambda_1}}{11520 }\frac{(1-\lambda_2)^{3/2}}{a^3} \left[7\pi^2 - \frac{60  (\mu a)^2}{1-\lambda_2} \right] .
\end{aligned}
\end{eqnarray}
And consequently, the Casimir pressure is given by
\begin{eqnarray}
\begin{aligned}
    P_{C} = \frac{\sqrt{1+\lambda_1}}{3840} \frac{(1-\lambda_2)^{3/2}}{a^4}\left[7\pi^2 - \frac{20(a\mu)^2}{1-\lambda_2} \right] .
\end{aligned}
\end{eqnarray}
\end{itemize}

\begin{figure}[h!]
	\centering
	\subfigure[Firt model.]{\includegraphics[width=0.4\textwidth]{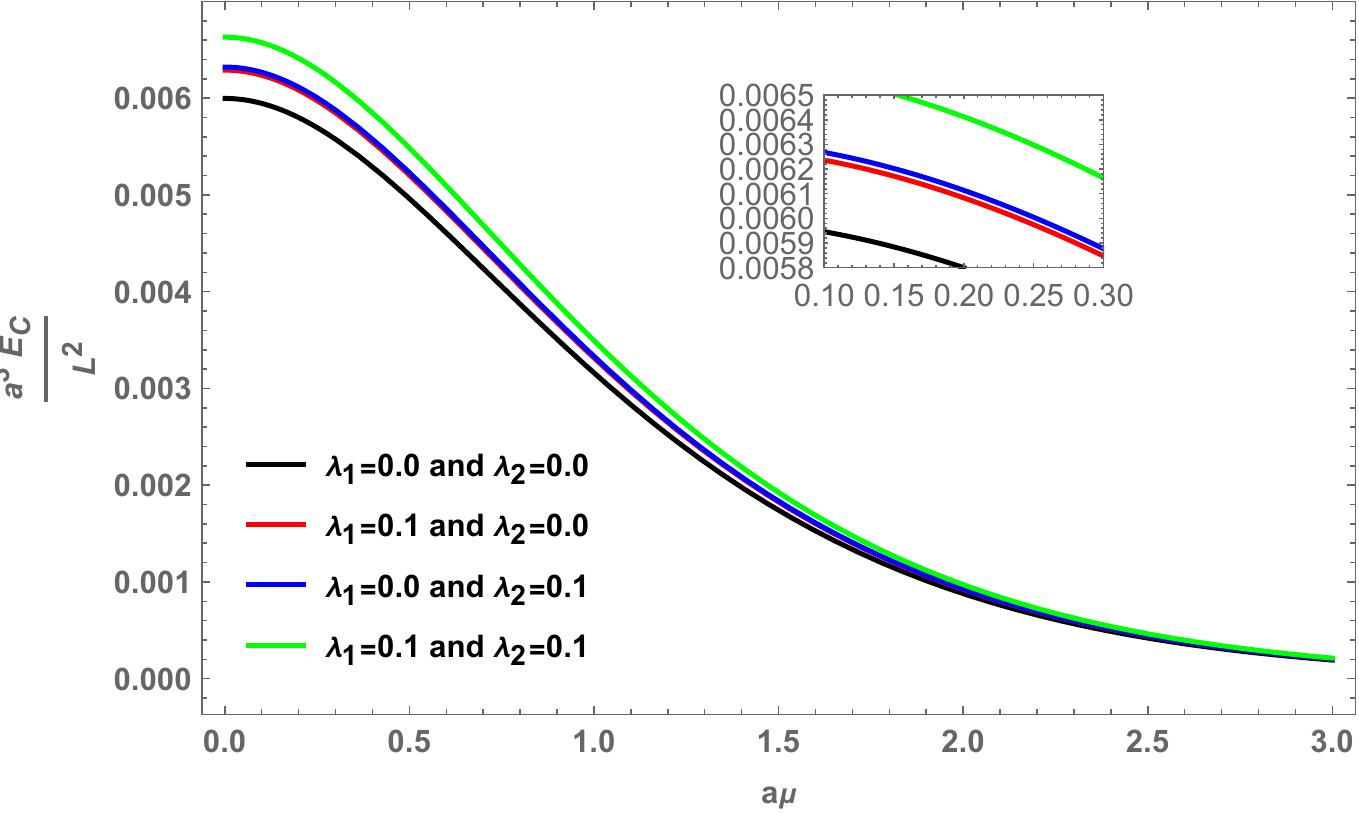}} 
	\subfigure[Second model.]{\includegraphics[width=0.4\textwidth]{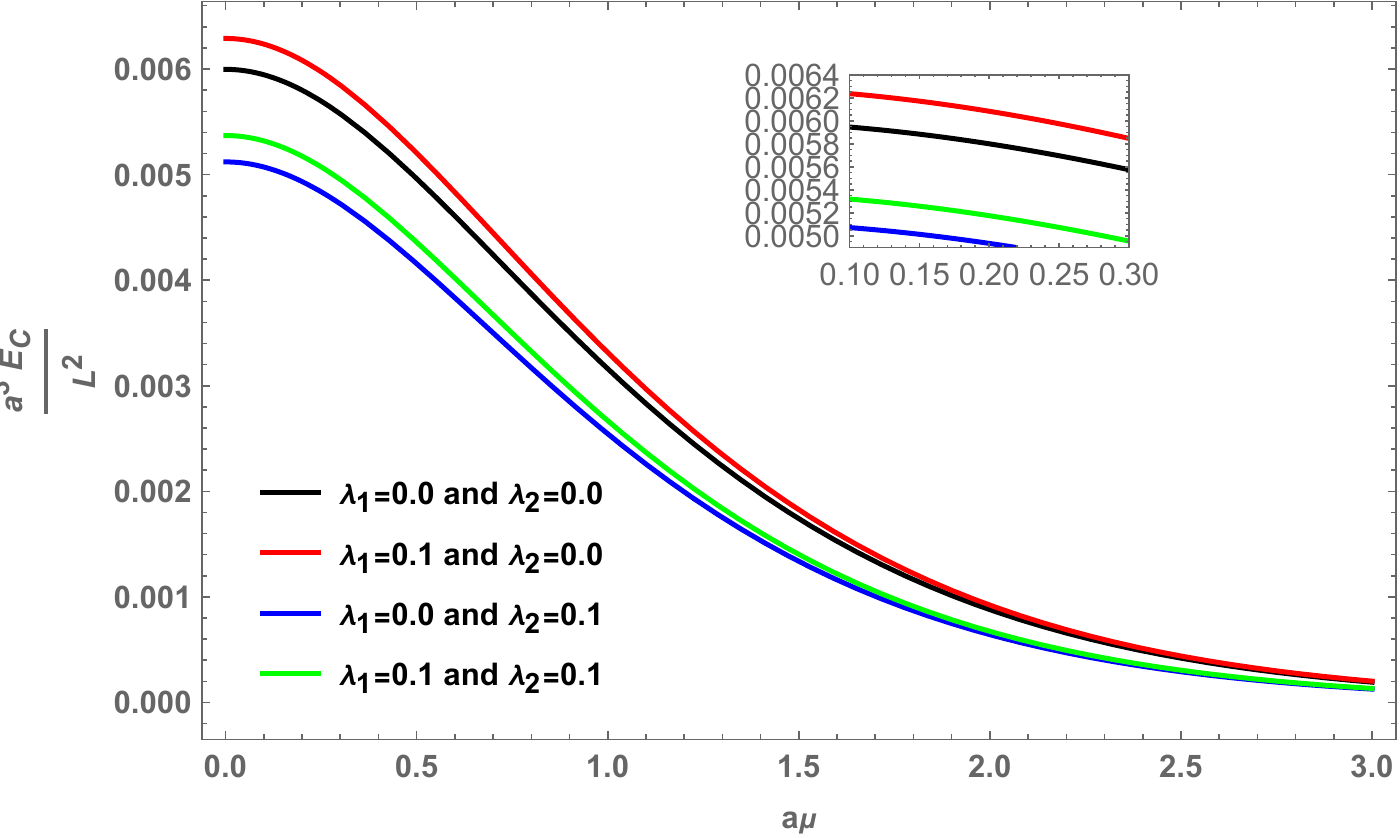}} 
	\caption{The behaviors of the Casimir energies per unit area multiplied by $a^3$ as a function of $a\mu$ are presented in the graph, adopting different values for the parameters $\lambda_1$ and $\lambda_2$. The left plot corresponds to the behavior of the four-vector $v^\mu$ parallel to the plates, while the right plot corresponds to the vector $v^\mu$ perpendicular to them.}
	\label{plot_mixed}
\end{figure}

As we can see, the results obtained in this section differ from the ones found for the Dirichlet and Neumann boundary conditions by a numerical factor and an opposite sign. These results corroborate the fact that the Casimir energy and pressure depend strongly on the boundary conditions imposed on the field and on parameters of Lorentz symmetry breaking. In Fig. \ref{plot_mixed}, we numerically display the behaviors of the Casimir energies per unit area as a function of $a\mu$, adopting different values for the parameters $\lambda_1$ and $\lambda_2$ for the two models analyzed in this section. The left plot shows the behavior associated with $v^\mu$ parallel to the plates, and the right plot shows the behavior for the vector $v^\mu$ perpendicular to them.

\section{Concluding remarks}
\label{Conclusions}

In this paper, we have analyzed the Casimir effect of a massive scalar quantum field confined between two large parallel plates in a Lorentz-violating scenario. The confinement of the bosonic field is achieved through boundary conditions imposed on the plates, namely Dirichlet, Neumann, and mixed conditions. Lorentz violation is introduced through a CPT-even, aether-like manner, considering the coupling between the space-time derivative of the field with two background constant four-vectors. The intensity of these interactions is determined by two dimensionless parameters, $\lambda_i$, for $i=1,2$. This analysis can be seen as an extension of the previous work developed in \cite{Messias2017}, where only one four-vector was considered. Here, we assume that one of the four-vectors is timelike and the other is spacelike. We consider two distinct directions for the latter: parallel and perpendicular to the plates.

Our results show that the Dirichlet and Neumann boundary conditions provide the same Casimir energies. These energies are expressed in integral representations, Eq. \eqref{int1}, for the second vector parallel to the plates, and by Eq. \eqref{int3a} for the vector perpendicular. Both integral representations can be expressed in terms of infinite series of Bessel functions, Eq. \eqref{int2} and Eq. \eqref{int4a}, respectively. For the second vector parallel to the plates, the modification in the Casimir energy due to the Lorentz violation is through a multiplicative factor; however, for the case perpendicular, the modification is more delicate. The correction associated with the timelike vector is through a multiplicative factor to the standard Casimir energy; however, for the spacelike vector, the modification can be understood as a situation where the distance between the plates is corrected by the factor $1/\sqrt{1-\lambda_2}$, as shown in Eq. \eqref{b-factor}. Using the integral representations for the Casimir energies per unit area, Eqs. \eqref{int1} and \eqref{int3a}, approximate expressions for both results can be obtained for $a\mu\ll1$ and $a\mu\gg1$. These results are given in Eq. \eqref{int3} and Eq. \eqref{appr_1}, for the vector parallel to the plates, and in Eq. \eqref{int4b} and Eq. \eqref{int4c}, for the vector perpendicular.

To better understand the influence of Lorentz violation on the Casimir energy, we plot the behavior of $a^3E/L^2$ as a function of $\mu a$ for both models analyzed, considering different values of the parameters $\lambda_1$ and $\lambda_2$: both parameters being zero (absence of Lorentz violation), only one of them being nonzero, and both parameters being nonzero. The corresponding graphs are given in Fig. \ref{plot_dirichlet}. From these plots, we can see that the presence of Lorentz violation presents corrections compared to the case without violation.

Another boundary condition imposed on the field on the plates was the mixed one, which was analyzed in section \ref{Mixed}. The configurations of the two four-vectors associated with LV remained the same: time-like and space-like. For the space-like vector parallel to the plates, an exact integral representation for the Casimir energy was provided in \eqref{casimir_energy_exact_mixed_1}. An infinite series expansion in terms of Bessel functions was also presented in Eq. \eqref{casimir_energy_exact_mixed_1a} for the corresponding Casimir energy. As we can see for this case, the Casimir energy is positive and the modification due to LV is through a multiplicative factor. Using the integral representation, approximate results for $E_C$ can be obtained. For $a \mu \gg 1$, we have Eq. \eqref{Mixed_1}, and for $a\mu \ll 1$, Eq. \eqref{Mixed_2}.

For the case where the space-like vector is orthogonal to the plates, the modification in the Casimir energy is more delicate, as explained in the previous paragraph. The integral representation for this energy is given in Eq. \eqref{casmir_energy_mixed_2}, which, on the other hand, can be expressed in terms of an infinite sum of modified Bessel functions, as shown in Eq. \eqref{casimir_energy_exact_mixed_1b}. Finally, approximate expressions are presented for $a\mu \gg 1$ in Eq. \eqref{Mixed_3}, and for $a\mu \ll 1$ in Eq. \eqref{casmir_energy_mixed_2_small}. Also, for this mixed boundary condition, graphs exhibiting the behavior of the Casimir energy per unit area as a function of $a\mu$ are presented, considering different values for $\lambda_1$ and $\lambda_2$. They are given in Fig. \ref{plot_mixed}, taking into account $\lambda_1=\lambda_2=0$, only $\lambda_2$ or $\lambda_1$ vanishing, and finally both parameters different from zero. The plots evidence the relevance of LV in the Casimir energy.

As our final comments, we would like to mention that another possible choice for the four-vector responsible for the LV could be a light-like one. However, in this case, the dispersion relation would not provide expressions for the Casimir energy that are very enlightening. For this reason, in this present investigation, we decided to consider two orthogonal four-vectors, one of them being time-like.  Another possible continuation of this present analysis is considering that the bosonic field satisfies the Robin boundary condition on the plates. For this case the momentum perpendicular to the plates, $k_z$, satisfies a transcendental equation involving  sine and cosine of $k_za$. Consequently, in order to perform a sum over the corresponding quantum number, a specific variant of the Abel-Plana summation formula is needed \cite{Romeo}. Because this topic is out of the scope of this paper, we decided do not included this investigation here. However it is our intention to develop this investigation in the near future.

\section*{Acknowledgments}

ERBM is partially supported by Conselho Nacional de Desenvolvimento Científico e Tecnológico (CNPq) under Grant No. 301.783/2019-3.

\end{document}